\documentclass[%
 reprint,
 amsmath,amssymb,
 aps,
pre,
showpacs,
nofootinbib
]{revtex4-1}

\usepackage{graphicx}
\usepackage{dcolumn}
\usepackage{bm}
\usepackage{natbib}

\begin{document}


\title{A Tight-Binding Approach to Overdamped Brownian Motion \\ on a Multidimensional Tilted Periodic Potential}

\author{K.~J.~Challis}
\author{Michael W.~Jack}%
\affiliation{%
 Scion, 49 Sala Street, Rotorua 3046, New Zealand
}%




\date{\today}

\begin{abstract}
We present a theoretical treatment of overdamped Brownian motion on a multidimensional tilted periodic potential that is analogous to the tight-binding model of quantum mechanics.  In our approach we expand the continuous Smoluchowski equation in the localized Wannier states of the periodic potential to derive a discrete master equation.  This master equation can be interpreted in terms of hopping within and between Bloch bands and for weak tilting and long times we show that a single-band description is valid.  In the limit of deep potential wells, we derive a simple functional dependence of the hopping rates and the lowest band eigenvalues on the tilt.  We also provide general expressions for the drift and diffusion in terms of the lowest band eigenvalues.

\end{abstract}


\maketitle


\section{\label{sec:intro} Introduction}

Brownian motion on a periodic potential has been used to describe a wide range of nonequilibrium transport phenomena where there are significant thermal fluctuations \cite{Risken89, Reimann02}.  For example, this arises in Josephson junction arrays \cite{Sterck02}, surface diffusion \cite{AlaNissila02}, particle sorting experiments \cite{Lacasta05, Astumian09, Xiao10}, and the mechanochemical coupling of molecular motors \cite{Magnasco94, Astumian97, Keller00, Golubeva12, VandenBroeck12, Challis12}.  The standard theoretical description of these systems is based on a continuous diffusion equation that governs the dynamics of the probability density and can be used to determine the average of macroscopic observables \cite{Risken89}.  In the limit of deep potential wells, it is physically intuitive that this continuous diffusion equation can be approximated by a discrete equation describing infrequent hopping transitions between potential wells.  Discrete treatments have been employed in the theory of chemical reactions \cite{Hanggi90, Gardiner85} and molecular motors \cite{Keller00,Fisher99,Lipowsky00, Golubeva12} and represent a significant simplification of the system dynamics.  In this paper we begin with the continuous diffusion equation and systematically transform to a discrete master equation for the case of overdamped Brownian motion on a multidimensional tilted periodic potential.

There have been previous attempts to discretize the continuous diffusion equation for Brownian motion on a tilted periodic potential.  In these treatments each cell of the periodic potential is identified with a discrete state and a master equation describes intercell hopping.  It has been suggested that these discrete equations can be made consistent with the continuous theory by matching the dynamic structure factor \cite{Ferrando93, Caratti98}, the eigenvalues \cite{Jung95}, or the ratio between backward and forward hopping \cite{Lattanzi01}.  Alternatively, a discrete model can be derived by connecting each spatial region via appropriate absorbing boundary conditions and source terms \cite{Lindner01}, or by  integrating the continuous probability density over a specific spatial region in the vicinity of each cell \cite{Keller00,Wang03a,Xing05}.  In contrast to these treatments, we implement the classical analog of the tight-binding model of a quantum particle in a periodic potential \cite{Kohn59, Jaksch98, Alfimov02, Kittel04}.  This is achieved by expanding the probability density of the continuous diffusion equation in a complete basis of Wannier states.  In the limit of deep potential wells these states are localized in the minima of the periodic potential and provide a natural basis for the system.  Using this basis we derive a discrete master equation that for deep potential wells is consistent with the physically-intuitive interpretation of hopping between wells.

In the multidimensional case, Brownian motion on a tilted periodic potential is not tractable analytically in general and instead has been investigated numerically \cite{Caratti97, Kostur00, Ghosh00, Wang03b}.  For non-separable potentials the tilt in one degree of freedom can induce a current in another and this coupling has important consequences in particle sorting experiments and molecular motors \cite{Magnasco94, Kostur00, Challis12}.  The tight-binding approach presented in this paper provides an analytic tool for treating multidimensional non-separable tilted periodic potentials.

This paper is organized as follows.  In section \ref{sec:norm} we introduce the Smoluchowski equation in our notation.  In section \ref{sec:Bloch} we expand the probability density for the system in the Bloch eigenfunction basis for the periodic potential and consider the time evolution of the Bloch bands.  In section \ref{sec:tightb} we transform to the  localized Wannier states of the periodic potential and derive a discrete master equation for the system.  In section \ref{sec:Kramers} we consider the limit of deep potential wells where the tilt dependence of the hopping rates can be determined analytically and hopping transitions occur predominantly between nearest neighbor wells.  In section \ref{sec:prop} we derive general expressions for the drift and diffusion.  We conclude in section \ref{sec:conc}.

\section{Smoluchowski equation \label{sec:norm}}

We consider Brownian motion of a particle in the overdamped limit of negligible inertia as described by the Smoluchowski equation \cite{Risken89, Gardiner85}
\begin{equation}
\frac{\partial P({\bm r},t)}{\partial t}= {\cal L} P({\bm r},t),
\label{Smoluchowskieqn}
\end{equation}
where $P({\bm r},t)$ is the probability density of finding the particle at position ${\bm r}$ at time $t$.  The evolution operator for the Smoluchowski equation is 
\begin{equation}
{\cal L} = \sum_{j}\frac{1}{ \gamma_j}\frac{\partial}{\partial r_j} \left[ \Theta \frac{\partial}{\partial r_j} + \frac{\partial V({\bm r})}{\partial r_j}\right] ,
\label{Lop}
\end{equation}
where $V({\bm r})$ is the external potential, $\Theta = k_B T$, $k_B$ is the Boltzmann constant, $T$ is the temperature, and $j$ is the coordinate index.  The friction coefficient ${\bm \gamma} $ may be different for each degree of freedom.

The external potential $V({\bm r})$ is the tilted periodic potential 
\begin{equation}
V({\bm r})= V_{\bm 0} ({\bm r})-{\bm f}\cdot {\bm r},
\label{pot_tiltper}
\end{equation}
where $V_{\bm 0} ({\bm r})$ is periodic with period ${\bm a}$, i.e.,
\begin{equation}
V_{\bm 0}({\bm r})= V _{\bm 0}({\bm r}+a_j\hat{{\bm r}}_j)= V_{\bm 0}({\bm r}+{\bm a}),
\end{equation}
and the linear potential represents a constant macroscopic force that drives the system out of thermal equilibrium and induces transport through the periodic potential.  We impose periodic boundary conditions with a period given by ${\bm N \bm a}$ where ${\bm N}$ is a diagonal matrix of integers.  We consider an infinite spatial extent by taking the limit $N_{jj}\rightarrow \infty$, although a similar approach is possible for $N_{jj}$ finite.  

\section{Bloch Eigenfunction Expansion  \label{sec:Bloch}}

The Smoluchowski equation (\ref{Smoluchowskieqn}) has separable solutions
\begin{equation}
P_{{\bm k}}(\bm r,t) = \phi_{\bm k} (\bm r) e^{-\lambda_{\bm k} t},
\end{equation} 
where the eigenfunctions $\phi_{{\bm k}}({\bm r})$ and eigenvalues $\lambda_{\bm k}$ satisfy the eigenequation
\begin{equation}
{\cal L}\phi_{\bm k}({\bm r})=-\lambda_{\bm k}\phi_{\bm k}(\bm r).
\label{Lestates}
\end{equation}
Due to the periodicity of ${\cal L}$, the eigenfunctions $\phi_{\bm k}(\bm r)$ can be chosen with the Bloch form \cite{Risken89, Kittel04}
\begin{equation}
{ \phi}_{\alpha, {\bm k}}({\bm r}) = e^{i {\bm k} \cdot {\bm r}} u_{\alpha,{\bm k}}({\bm r}),
\label{phi_Bloch}
\end{equation}
where $u_{\alpha,{\bm k}}({\bm r})$  has the same periodicity as the potential $V_{\bm 0}({\bm r})$, $\alpha$ is the band index, and ${\bm k}$ is continuous\footnote{For a finite spatial extent, $N_{jj}$ is finite and the label ${\bm k}$ is quantized.} and restricted to the first Brillouin zone, i.e.,
\begin{equation}
-\frac{\pi}{a_j} \leq k_j \leq \frac{\pi}{a_j}.
\label{FBZ}
\end{equation}

The evolution operator ${\cal L}$ can be transformed to the self-adjoint operator \cite{Risken89}
\begin{eqnarray}
{\cal H} & = &  e^{V_{\bm f}({\bm r})/2\Theta} {\cal L} e^{-V_{\bm f}({\bm r})/2\Theta}  \label{Herm_op_v1} \\
& =& \sum_{j}\frac{1}{\gamma_j}\left[ \Theta  \frac{\partial^2}{\partial r_j^2} -U_j({\bm r}) \right],
\label{Herm_op}
\end{eqnarray}
where
\begin{equation}
U_j({\bm r}) = \frac{1}{4 \Theta} \left[ \frac{\partial V({\bm r})}{\partial r_j}\right]^2-\frac{1}{2} \frac{\partial ^2 V({\bm r})}{\partial r_j^2}.
\label{Upot}
\end{equation}
The operator ${\cal H}$ is not Hermitian in general.  However, denoting quantities for the ${\bm f} = {\bm 0}$ case by a zero subscript (or superscript), the operator ${\cal H}_{\bm 0}$ is Hermitian and the untilted Bloch eigenfunctions ${ \phi}^{\bm 0}_{\alpha, {\bm k}}({\bm r})$ form a complete orthonormal basis (see Appendix \ref{app:orthonorm}).  The untilted Bloch eigenfunctions satisfy the orthonormality relation
\begin{equation}
\int d{\bm r} \ e^{V_{\bm 0}({\bm r})/ \Theta} \phi_{\alpha,{\bm k}}^{\bm 0*}({\bm r}) \phi^{\bm 0}_{\alpha',{\bm k}'}({\bm r}) = \delta_{\alpha \alpha'}\delta({\bm k}-{\bm k}'),
\end{equation}
and the completeness relation
\begin{equation}
e^{ V_{\bm 0}({\bm r}) / \Theta} \sum_{ \alpha}\int_{\cal B} d{\bm k} \   \phi_{\alpha,{\bm k}}^{\bm 0*} ({\bm r}) \phi^{\bm 0}_{\alpha, {\bm k}} ({\bm r}') = \delta ({\bm r}-{\bm r}'),
\label{complete_phi0}
\end{equation}
where the integral in Eq.\ (\ref{complete_phi0}) is denoted by ${\cal B}$ to indicate that it is over a single Brillouin zone [see Eq.\ (\ref{FBZ})].  The eigenvalues $\lambda_{\alpha,\bm k}^{\bm 0}$ are real and exist in bands separated by forbidden band gaps (see Appendix \ref{app:orthonorm}).

Smoluchowski equation solutions satisfying periodic boundary conditions at infinity can be expanded without loss of generality in the untilted Bloch eigenfunction basis.  The expanded probability density is
\begin{equation}
P({\bm r},t) = \sum_{\alpha} \int_{\cal B} d{\bm k} \ c_{\alpha, {\bm k}} (t) \phi_{\alpha,{\bm k}}^{\bm 0} ({\bm r}),
\label{Pexpc}
\end{equation}
where 
\begin{equation}
c_{\alpha, \bm k}(t)= \int d\bm r \ e^{V_{\bm 0}(\bm r)/ \Theta}  \phi_{\alpha,\bm k}^{\bm 0*} P(\bm r,t)  .
\end{equation}
Transforming the Smoluchowski equation (\ref{Smoluchowskieqn}), the coefficients $c_{\alpha,\bm k}(t)$ evolve according to
\begin{equation}
\frac{d c_{\alpha,\bm k}(t)}{dt} = \sum_{\alpha'}\int _{\cal B} d \bm k' \ \mu_{\alpha,\alpha',\bm k, \bm k'} c_{\alpha',\bm k'}(t) ,
\label{dcdt}
\end{equation}
where 
\begin{equation}
\mu_{\alpha,\alpha',\bm k, \bm k'} = \int d\bm r \ e^{V_{\bm 0}(\bm r)/ \Theta} {\phi}^{\bm 0*}_{\alpha,\bm k} ({\bm r}){\cal L} {\phi}^{\bm 0}_{\alpha',{\bm k}'} ({\bm r}).
\label{mu}
\end{equation}

In the $\bm f = \bm 0 $ case, 
\begin{equation}
\mu^{\bm 0}_{\alpha,\alpha',\bm k,\bm k'} = -\lambda_{\alpha,\bm k}^{\bm 0} \delta_{\alpha \alpha'} \delta (\bm k-\bm k'),
\end{equation}
and the evolution equation (\ref{dcdt}) becomes 
\begin{equation}
\frac{d c^{\bm 0}_{\alpha,\bm k}(t)}{dt} = -\lambda^{\bm 0}_{\alpha,\bm k} c^{\bm 0}_{\alpha,\bm k}(t).
\label{master_diag_0}
\end{equation} 
Equation (\ref{master_diag_0}) can be integrated directly to give
\begin{equation}
c^{\bm 0}_{\alpha,\bm k}(t) = c^{\bm 0}_{\alpha,\bm k}(0) e^{-\lambda_{\alpha,\bm k}^{\bm 0} t}. 
\label{diag_sol_0}
\end{equation}
Equation (\ref{diag_sol_0}) shows that each eigenfunction coefficient $c^{\bm 0}_{\alpha,\bm k}(t)$ decays in time with a rate given by its eigenvalue $\lambda_{\alpha,\bm k}^{\bm 0} \geq 0$ (see Appendix \ref{app:pose}).  The band gap between the lowest and first Bloch bands enables a separation of time scales so that for $t \gg 1/\min(\lambda_{1,\bm k}^{\bm 0})$ all bands with $\alpha>0$ are damped out and only the lowest $\alpha=0$ band remains occupied.  

In the tilted periodic potential,
\begin{equation}
\mu_{\alpha,\alpha',\bm k,\bm k'} = -\lambda_{\alpha,\bm k}^{\bm 0} \delta_{\alpha \alpha'}\delta (\bm k-\bm k')-\nu_{\alpha,\alpha',\bm k}\delta(\bm k-\bm k'),
\label{mu_eqn}
\end{equation}
and the evolution equation (\ref{dcdt}) becomes 
\begin{equation}
\frac{d c_{\alpha,\bm k}(t)}{dt} = -\lambda^{\bm 0}_{\alpha,\bm k} c_{\alpha,\bm k}(t)-\sum_{\alpha'}\nu_{\alpha,\alpha',\bm k} c_{\alpha',\bm k}(t).
\label{master_diag}
\end{equation} 
The first term on the right-hand side of Eq.\ (\ref{master_diag}) is due to the untilted periodic potential $V_{\bm 0}({\bm r})$ and leads to decay of the coefficients $c_{\alpha,\bm k}(t)$ at the rate $\lambda_{\alpha,\bm k}^{\bm 0} $ [see Eq.\ (\ref{diag_sol_0})].  The second term is due to the linear potential and couples different Bloch bands of the untilted periodic potential.  If the tilt ${\bm f}$ is sufficiently small that the band gaps in the decay rates $\lambda^{\bm 0}_{\alpha,\bm k}$ are large compared to the coupling $\nu _{\alpha,\alpha',\bm k}$, i.e.,
\begin{equation}
\left|\nu _{\alpha,\alpha',\bm k} \right| \ll \left|\lambda^{\bm 0}_{\alpha,\bm k}-\lambda^{\bm 0}_{\alpha',\bm k}\right|,
\label{criterion}
\end{equation} 
interband coupling can be neglected.  In this \emph{weak-tilting regime}, the evolution equation (\ref{master_diag}) can be approximated by
\begin{eqnarray}
\frac{d c_{\alpha,\bm k}(t)}{dt}&  =&  -[\lambda^{\bm 0}_{\alpha,\bm k} +\nu_{\alpha,\alpha,\bm k}] c_{\alpha,\bm k}(t). \label{master_diag_wt}
\end{eqnarray}
Equation (\ref{master_diag_wt}) can be integrated analytically and reduces to Eq.\ (\ref{master_diag_0}) for $\bm f = \bm 0$ where $\nu_{\alpha,\alpha,\bm k}=0$.  

\section{Wannier Basis \label{sec:tightb}}

In the case where the periodic potential $V_{\bm 0} ({\bm r})$ creates deep potential wells that localize the particle, the untilted Bloch eigenfunctions $\phi^{\bm 0}_{\alpha,\bm k} ({\bm r})$ are not a natural basis because they are delocalized over the entire spatial extent of the system.  It is more convenient to transform to the Wannier states that are localized around the potential wells of the periodic potential.  The Wannier states for the eigenfunctions ${ \phi}^{\bm 0}_{\alpha,\bm k}({\bm r})$ are
\begin{equation}
w_{\alpha,{\bm n}} ({\bm r})= D \int_{\cal B} d{\bm k} \ {\phi}^{\bm 0}_{\alpha,\bm k} ({\bm r})e^{-i {\bm k} \cdot {\bm A}  {\bm n}},
\label{wstates}
\end{equation}
 where ${\bm A}$ is a diagonal matrix with $A_{jj}= a_j$, the constant $D$ is
\begin{equation}
D = \prod_{j} \left( \frac{a_j}{2\pi} \right),
\end{equation}
and for brevity we have omitted the zero superscript on the untilted Wannier states (\ref{wstates}).  Taking the complex conjugate of the eigenequation (\ref{Lestates}) with ${\bm f} = \bm 0$, the untilted Bloch eigenfunctions are found to satisfy $\phi_{\alpha,\bm k}^{\bm 0*}({\bm r})=\phi^{\bm 0}_{\alpha,-\bm k}({\bm r})$ so the Wannier states (\ref{wstates}) are real.  The untilted Wannier states form a complete orthonormal basis.  They satisfy the orthonormality relation
\begin{equation}
\frac{1}{D} \int d{\bm r} \ e^{V_{\bm 0}({\bm r})/\Theta} w_{\alpha,{\bm n}} ({\bm r})w_{\alpha',\bm n'} ({\bm r}) =  \delta_{\alpha \alpha'} \delta_{{\bm n} {\bm n}'},
\end{equation}
and the completeness relation
\begin{equation}
\frac{1}{D} e^{V_{\bm 0}({\bm r})/\Theta}\sum_{\alpha,\bm n} w_{\alpha, {\bm n}} ({\bm r}) w_{\alpha, {\bm n}} ({\bm r}') =  \delta ({\bm r}-{\bm r}').
\label{complete}
\end{equation}

The probability density $P(\bm r,t)$ can be expanded in the untilted Wannier states as
\begin{equation}
P({\bm r},t) = \frac{1}{D} \sum_{\alpha,\bm n} p_{\alpha,\bm n}(t) w_{\alpha,\bm n}({\bm r}),
\label{Pexpw}
\end{equation}
where
\begin{equation}
p_{\alpha,\bm n}(t) = \int d{\bm r} \ e^{V_{\bm 0}({\bm r})/\Theta} w_{\alpha,\bm n}({\bm r}) P({\bm r},t) .
\label{pncoeff}
\end{equation}
The coefficients $p_{\alpha,\bm n}(t)$ are real and their amplitude can be interpreted as the probability that the Brownian particle occupies the Wannier state $w_{\alpha,\bm n}({\bm r})$.  Transforming the Smoluchowski equation (\ref{Smoluchowskieqn}), the evolution of the system can be described by the master equation \cite{Gardiner85}
\begin{equation}
\frac{d p_{\alpha,\bm n}(t)}{d t} = \sum_{\alpha',{\bm n}'} \sigma_{\alpha,\alpha', {\bm n},{\bm n}'} p_{\alpha',{\bm n}'}(t),
\label{pneqn}
\end{equation}
where
\begin{eqnarray}
\sigma_{\alpha,\alpha', {\bm n},{\bm n}'}  =  \frac{1}{D} \int d{\bm r} \ e^{V_{0}({\bm r})/\Theta}w_{\alpha,\bm n}({\bm r})  {\cal L}  w_{\alpha',{\bm n}'}  ({\bm r}).
\label{sigma}
\end{eqnarray}
The \emph{coupling matrix} defined by Eq.\ (\ref{sigma}) is real and, recognising that the Wannier states satisfy
\begin{equation}
w_{\alpha,\bm n}({\bm r})= w_{\alpha,{\bm 0}}({\bm r}-{\bm A} {\bm n}),
\label{wna}
\end{equation}
we find that
\begin{equation}
\sigma_{\alpha,\alpha', {\bm n},{\bm n}'} = \sigma_{\alpha,\alpha',{\bm n}-{\bm n}',{\bm 0}} = \sigma_{\alpha,\alpha', {\bm 0},{\bm n}'-\bm n}.
\end{equation}

In the $\bm f = \bm 0 $ case, the coupling matrix becomes
\begin{eqnarray}
\sigma^{\bm 0}_{\alpha,\alpha',{\bm n},{\bm n}'} & = & \frac{1}{D} \int d{\bm r} \ e^{V_{\bm 0}({\bm r})/\Theta}w_{\alpha,\bm n}({\bm r})  {\cal L}_{\bm 0}  w_{\alpha',{\bm n}'}  ({\bm r}) \\
& =&  \kappa^{\bm 0}_{\alpha,{\bm n}-\bm n'}\delta_{\alpha\alpha'},
\end{eqnarray}
where $\kappa^{\bm 0}_{\alpha,\bm n}$ are the Fourier components of the ${\cal L}_{\bm 0}$ eigenvalues $\lambda_{\alpha,\bm k}^{\bm 0}$, i.e.,
\begin{equation}
\kappa^{\bm 0}_{\alpha,\bm n} =- D\int_{\cal B}d{\bm k} \ \lambda_{\alpha,\bm k}^{\bm 0}e^{i {\bm k}\cdot {\bm A \bm n}} .
\label{kappa0}
\end{equation}
The master equation (\ref{pneqn}) becomes
\begin{equation}
\frac{d p_{\alpha,\bm n}^{\bm 0}(t)}{d t} = \sum_{{\bm n}'} \kappa^{\bm 0}_{\alpha, {\bm n}-\bm n'} p^{\bm 0}_{\alpha, {\bm n}'}(t),
\label{pneqnv2}
\end{equation}
and each band $\alpha$ evolves independently.  The eigenvalues $\lambda_{\alpha,\bm k}^{\bm 0}$ are real (see Appendix \ref{app:orthonorm}) and symmetric in ${\bm k}$ (due to the Bloch form), so by Eq.\ (\ref{kappa0}) the hopping rates $\kappa_{\alpha,\bm n}^{\bm 0}$ are real and symmetric in ${\bm n}$, i.e.,
\begin{equation}
\kappa^{\bm 0}_{\alpha,\bm n}=\kappa^{\bm 0}_{\alpha,-\bm n}=\kappa^{\bm 0*}_{\alpha,\bm n}.
\label{kappa_sym}
\end{equation} 
Equation (\ref{kappa_sym}) shows that the forward and backward rates of thermally activated hopping transitions between potential wells are equal.  This is the condition of detailed balance  \cite{Gardiner85}.

In the tilted periodic potential, the coupling matrix is
\begin{eqnarray}
\sigma_{\alpha,\alpha',{\bm n},{\bm n}'} & = &\kappa^{\bm 0}_{\alpha, {\bm n}-\bm n '} \delta_{\alpha \alpha'} + \Delta_{\alpha,\alpha',{\bm n}-\bm n'} \label{sigma_split}\\
& = & \kappa_{\alpha,{\bm n}-\bm n'} \delta_{\alpha\alpha'} + \Delta_{\alpha,\alpha',{\bm n}-\bm n'} (1-\delta_{\alpha\alpha'}),
\label{sigma_v2}
\end{eqnarray}
and the master equation (\ref{pneqn}) becomes
\begin{equation}
\frac{d p_{\alpha,\bm n}(t)}{d t} =  \sum_{\bm n'}  \kappa^{\bm 0}_{\alpha,{\bm n}-\bm n'} p_{\alpha,{\bm n}'}(t)+  \sum_{\alpha',{\bm n}'} \Delta_{\alpha,\alpha',\bm n-\bm n'} p_{\alpha',{\bm n}'}(t).
\label{pneqnv3}
\end{equation}
The contribution $\Delta_{\alpha,\alpha',{\bm n}}$ from the linear potential both modifies the intraband hopping rates and gives rise to coupling between the untilted Bloch bands.  For strongly tilted periodic potentials the potential wells created by the periodic potential $V_{\bm 0}(\bm r)$ are substantially modified by the tilt and coupling between the untilted Bloch bands may be significant.  The strong tilting regime has been considered previously in both one   \cite{Reimann01, Monnai07, SalgadoGarcia08} and two dimensions \cite{Kostur00}.

\subsection{Weak-Tilting Regime \label{sec:weak}}
 
The Wannier basis is particularly useful when the force $\bm f$ is not strong enough to significiantly modify the potential wells of the periodic potential $V_{\bm 0}(\bm r)$ and interband coupling is negligible.  This weak-tilting regime is defined by Eq.\ (\ref{criterion}) and, while interband coupling is negligible, the linear potential modifies the intraband hopping rates.  In the weak-tilting regime, the master equation (\ref{pneqnv3}) can be well approximated by 
\begin{eqnarray}
\frac{d p_{\alpha,\bm n}(t)}{dt} & = & \sum_{{\bm n}'} \kappa_{\alpha, {\bm n}-\bm n'} p_{\alpha,{\bm n}'} (t),
\label{pneqn_alpha}
\end{eqnarray}
where the intraband hopping rates are
\begin{eqnarray}
\kappa_{\alpha,\bm n} & = & \frac{1}{D}\int d{\bm r} \ e^{V_{\bm 0}({\bm r})/\Theta}w_{\alpha,{\bm n}}({\bm r})  {\cal L}  w_{\alpha,\bm 0}  ({\bm r}) \\
& = & \kappa_{\alpha,\bm n}^{\bm 0} +\Delta_{\alpha,\alpha,\bm n}.
\label{kappa_int}
\end{eqnarray}
Neglecting interband coupling, the hopping rates $\kappa_{\alpha,\bm n}$ are the  Fourier components of the ${\cal L}$ eigenvalues $\lambda_{\alpha,\bm k}$ (see Appendix \ref{app:evalues_tilt}), i.e.,
\begin{equation}
\kappa_{\alpha,\bm n} \approx- D \int_{\cal B}d{\bm k} \  \lambda_{\alpha,\bm k}e^{i {\bm k}\cdot {\bm A \bm n}}.
\label{kappa}
\end{equation}
The intraband hopping rates are real, i.e.,
\begin{equation}
\kappa_{\alpha,\bm n}= \kappa^{*}_{\alpha,\bm n},
\end{equation}
but the linear potential breaks the symmetry so that $\lambda_{\alpha,\bm k}$ is imaginary in general (see Appendix \ref{app:orthonorm}) and $\kappa_{\alpha,\bm n} \neq \kappa_{\alpha,-\bm n}$.  In the $\bm f = \bm 0$ case, Eqs.\ (\ref{pneqn_alpha}) and (\ref{kappa}) become Eqs.\ (\ref{pneqnv2}) and (\ref{kappa0}), respectively. 

In the weak-tilting regime, coupling between the untilted Bloch bands is negligible and each band evolves independently.  For $t \lesssim 1/\lambda_{1, \bm 0}^{\bm 0}$, the system evolution is dominated by damping of the $\alpha\neq0$ Bloch bands, as described in section \ref{sec:Bloch}.  After this time the higher Bloch bands are barely occupied and the dynamics of the Brownian particle is dominated by intraband hopping within the $\alpha=0$ band.  This $\alpha=0$ band preserves normalization under the evolution operator $\cal L$ (see Appendix \ref{app:norm}).  Retaining only the stable $\alpha=0$ band and dropping the $\alpha$ subscript, the evolution of the Brownian particle can be described by the single-band master equation
\begin{equation}
\frac{d p_{{\bm n}}(t)}{d t} = \sum_{{\bm n}'} \kappa_{{\bm n}-{\bm n}'} p_{{\bm n}'}(t).
\label{master_eqn_F}
\end{equation}
Inverting Eq.\ (\ref{kappa}) gives the eigenvalues
\begin{equation}
\lambda_{\alpha,\bm k} \approx -\sum_{\bm n} \kappa_{\alpha,\bm n} e^{-i \bm k \cdot \bm A \bm n}.
\label{lambda_kappas}
\end{equation}
Taking $\bm k = \bm 0$ gives
\begin{equation}
\sum_{\bm n} \kappa_{\alpha,\bm n}\approx -\lambda_{\alpha,\bm 0},
\end{equation} 
and, using that $\lambda_{0,\bm 0}=0$ (see Appendix \ref{app:norm}), the master equation (\ref{master_eqn_F}) preserves normalization and can be cast into the more familiar form \cite{Gardiner85, Risken89}
\begin{equation}
\frac{d p_{{\bm n}}(t)}{d t}  =  \sum_{{\bm n}' } \left[ \kappa_{{\bm n}-{\bm n}'} p_{{\bm n}'}(t) -\kappa_{{\bm n}'-{\bm n}} p_{\bm n}(t) \right].
\label{master_eqn_F_v2}
\end{equation}
Equation (\ref{master_eqn_F_v2}) describes hopping between Wannier states within the $\alpha=0$ Bloch band.  With the hopping rates (\ref{kappa}), Eq.\ (\ref{master_eqn_F_v2}) is a single-band discrete master equation describing Brownian motion on a weakly-tilted periodic potential and is the key result of this paper.  

The master equation (\ref{master_eqn_F_v2}) represents a significant simplification of the system dynamics and can be solved by transforming to the diagonal form
\begin{equation}
\frac{d c_{\bm k}(t)}{dt} = -\lambda_{\bm k} c_{\bm k}(t),
\label{discrete_diag}
\end{equation}
where the eigenstates 
\begin{equation}
c_{\bm k}(t) = \sum_{\bm n} p_{\bm n}(t) e^{-i \bm k \cdot \bm A \bm n},
\end{equation}
are the expansion coefficients of the probability density in the untilted Bloch eigenfunction basis [see Eq.\ (\ref{Pexpc})] and $\lambda_{\bm k}$ are the eigenvalues of the evolution operator ${\cal L}$ \cite{Jung95}, in the weak-tilting regime where interband coupling is negligible.  Equation (\ref{discrete_diag}) can be integrated analytically.

\subsection{Current}

The Smoluchowski equation (\ref{Smoluchowskieqn}) can be written as
\begin{equation}
\frac{\partial P(\bm r, t)}{\partial t} = -\nabla \cdot \bm J(\bm r,t),
\end{equation}
where the coordinates of the current density are
\begin{equation}
J_j (\bm r,t) = -\frac{1}{\gamma_j}\left[ \Theta \frac{\partial}{\partial r_j} + \frac{\partial V_{\bm f}(\bm r)}{\partial r_j} \right] P(\bm r,t).
\label{J}
\end{equation}
The current density can be expanded in the Wannier basis as
\begin{equation}
{\bm J}({\bm r},t) = \frac{1}{D} \sum_{\alpha,\bm n} {\bm j}_{\alpha,\bm n}(t) w_{\alpha,\bm n}({\bm r}),
\label{Jexp}
\end{equation}
with the expansion coefficients 
\begin{equation}
{\bm j}_{\alpha,\bm n}(t) = \int d{\bm r} \ e^{V_{\bm 0}({\bm r})/\Theta} w_{\alpha,\bm n}({\bm r}){\bm J}({\bm r},t).
\end{equation}
The current for the Brownian particle is
\begin{equation}
{\bm I}  =  \int  d{\bm r} \ {\bm J}({\bm r},t) = \sum_{\alpha,\bm n} {\bm j}_{\alpha,\bm n}(t) \xi_{\alpha}.
\label{current}
\end{equation}
Equation (\ref{current}) is equal to the particle drift (see Section \ref{sec:prop}) when the current density vanishes at the system boundary \cite{Reimann02}.

In the weak-tilting regime retaining only the $\alpha=0$ band, the current density expansion coefficients can be determined to be
\begin{eqnarray}
{\bm j}_{{\bm n}}(t) & = & -\frac{\Theta }{\bm \gamma} \sum_{{\bm n}'} p_{{\bm n}'}(t)\frac{1}{D} \int d{\bm r} \ w_{{\bm n}}({\bm r})\nabla e^{V_{\bm 0}({\bm r})/\Theta} w_{{\bm n}'}({\bm r}) \nonumber \\
& &  +\frac{\bm f}{\bm \gamma} p_{{\bm n}}(t),
\label{jnt}
\end{eqnarray}
where the vector division is taken elementwise.  Taking the sum over ${\bm n}$ and using the Bloch form for the eigenstates with the ground state $\phi_{0, \bm 0}^{\bm 0}(\bm r)\propto e^{-V_{\bm 0}(\bm r)/\Theta}$, the contribution to the current (\ref{current}) from the first term in Eq.\ (\ref{jnt}) can be shown to vanish.  Therefore,
\begin{equation}
{\bm I} = {\cal N}\frac{ \bm f}{\bm \gamma},
\label{I}
\end{equation}
i.e., for weak-tilting and long times, the steady-state current is proportional to the tilt.

\section{Tight-Binding Limit \label{sec:Kramers}}

We consider the \emph{tight-binding limit} where the Wannier states are strongly localized around a single period of the potential $V_{\bm 0}(\bm r)$ and the linear potential is sufficiently slowly varying that it changes negligibly across the spatial extent of any given Wannier state. With these conditions satisfied, it is possible to analytically derive the functional dependence of the hopping rates $\kappa_{\alpha,\bm n}$ and the eigenvalues $\lambda _{\alpha,\bm k}$ on the force $\bm f$ and to further explore the dynamic behavior of the system without specific knowledge of the periodic potential.

The validity criteria for the tight-binding limit can be determined formally as follows.  The Wannier states are well localized if the periodic potential has deep potential wells, i.e., when the amplitude of the periodic potential is large compared to the thermal energy $\Theta$. With this condition satisfied the shape of the Wannier states (particularly in the lowest Bloch band) depend predominantly on the curvature of the periodic potential at its extrema.  Approximating the periodic potential in the vicinity of its extrema by a harmonic potential (see Appendix \ref{app:tight_binding}), the linear potential is a small perturbation near these extrema when
\begin{equation}
|f_j| \ll a_j \left| \frac{\partial^2 V_{\bm 0}(\bm r)}{\partial r_j^2}\right|_{\bm r = \bm r_{\rm ext}},
\label{tight_binding}
\end{equation}
where ${\bm r}_{\rm ext}$ are the positions of extrema in the periodic potential.  Equation (\ref{tight_binding}) ensures that the tilt varies slowly across the extent of the Wannier states and that coupling between Bloch bands is negligible (see Section \ref{sec:weak}).

In general the coupling matrix $\sigma_{\alpha,\alpha', \bm n,\bm n'}$ of Eq.\ (\ref{sigma}) can be expressed in terms of the self-adjoint operator ${\cal H}$ [see Eq.\ (\ref{Herm_op_v1})] as
\begin{eqnarray}
\sigma_{\alpha,\alpha',\bm n,\bm n'} & = &  \frac{1}{D}\int d\bm r \ e^{V_{\bm 0}({\bm r})/\Theta} e^{-V({\bm r})/2\Theta}w_{{\alpha,\bm n}}({\bm r}) \nonumber \\
& &\quad \quad \times {\cal H} e^{V({\bm r})/2\Theta}  w_{\alpha',{\bm n}'}  ({\bm r}) .
\label{sigma_H}
\end{eqnarray}
To evaluate Eq.\ (\ref{sigma_H}) in the tight-binding limit we insert the completeness relation (\ref{complete}) and take the first term in the Taylor expansion of  ${\pm {\bm f}\cdot {\bm r}/2\Theta}$ around the position ${\bm A\bm n}$ of the appropriate Wannier state, i.e.,
\begin{eqnarray}
 \lefteqn{\frac{1}{D}\int d{\bm r} \ e^{V_{\bm 0}({\bm r})/\Theta} e^{\pm {\bm f}\cdot {\bm r}/2\Theta} w_{\alpha,\bm n} ({\bm r})  w_{\alpha',\bm n '} ({\bm r}) }\nonumber  \\
 & & \quad \quad \quad \quad \quad \quad \quad \quad \quad \approx e^{\pm {\bm f}\cdot {\bm A \bm n}/2\Theta} \delta_{\alpha \alpha'} \delta_{{\bm n} {\bm n}'}.
\end{eqnarray}
This gives
\begin{eqnarray}
\sigma_{\alpha,\alpha',\bm n,\bm n'} & \approx & e^{-\bm f \cdot \bm A(\bm n'-\bm n)/2\Theta} \frac{1}{D}\int d\bm r \ e^{V_{\bm 0}({\bm r})/2\Theta}w_{{\alpha,\bm n}}({\bm r}) \nonumber \\
& &    \times {\cal H} e^{V_{\bm 0}({\bm r})/2\Theta}  w_{\alpha',{\bm n}'}  ({\bm r}) .
\label{sigma_Hv2}
\end{eqnarray}
The dominant contribution to the integral (\ref{sigma_Hv2}) comes from the spatial regions around the extrema of the periodic potential where the Wannier states are finite and the potential is approximately harmonic \cite{Caroli79}.  In these regions the operator ${\cal H}$ can be approximated by ${\cal H}_{\bm 0}$ and
\begin{eqnarray}
\sigma_{\alpha,\alpha',\bm n,\bm n'} & \approx&  e^{-\bm f \cdot \bm A(\bm n'-\bm n)/2\Theta} \frac{1}{D}\int d\bm r \ e^{V_{\bm 0}({\bm r})/2\Theta} w_{\alpha,\bm n}(\bm r) \nonumber \\
& & \times {\cal H}_{\bm 0} e^{V_{\bm 0}({\bm r})/2\Theta}  w_{\alpha',{\bm n}'}  ({\bm r})  \\
& = &  \kappa_{\alpha,\bm n-\bm n'}\delta_{\alpha\alpha'},
\label{sigma_Hv3}
\end{eqnarray}
with
\begin{equation}
\kappa_{\alpha,\bm n} = e^{\bm f \cdot \bm A \bm n/2\Theta}\kappa_{\alpha,\bm n}^{\bm 0}.
\label{kappa_f} 
\end{equation}
Equation (\ref{sigma_Hv3}) shows that, in the tight-binding limit, the coupling matrix $\sigma_{\alpha,\alpha',\bm n,\bm n'}$ is diagonal in the band index and the bands decouple in the same way that they decouple in the weak-tilting regime (see Section \ref{sec:weak}).  Equation (\ref{kappa_f}) provides a simple functional dependence of the hopping rates on the tilt.  This can be used to derive the ratio of forward to backward hopping rates, i.e., 
\begin{equation}
\frac{\kappa_{\alpha,\bm n}}{\kappa_{\alpha,-\bm n} } = e^{\bm f \cdot \bm A\bm n/\Theta}.
\label{rate_ratio}
\end{equation}
Equation (\ref{rate_ratio}) is a generalized detailed balance condition for tilted periodic potentials \cite{Lattanzi01,Wang03a, Xing05, Astumian09}.  It is independent of the exact form of the periodic potential and the band index $\alpha$.  However, the Wannier states become increasingly delocalized with increasing band index so Eq.\ (\ref{rate_ratio}) can be expected to become less accurate with increasing $\alpha$.

In the tight-binding limit, the tilt dependence of the Bloch eigenvalues $\lambda_{\alpha,\bm k}$ can be determined by inserting the tilt dependence (\ref{kappa_f}) of the hopping rates into Eq.\ (\ref{lambda_kappas}). This yields 
\begin{equation}
\lambda_{\alpha,\bm k}  \approx - \sum_{\bm n} \kappa_{\alpha,\bm n}^{\bm 0} e^{\bm f \cdot \bm A \bm n/2\Theta} e^{-i \bm k \cdot \bm A\bm n}.
\label{lambda_Kramers}
\end{equation} 
In the case of tight-binding, nearest-neighbor hopping dominates \cite{Alfimov02}.  This means that higher-order hopping across multiple wells can be neglected and the summation in Eq.\ (\ref{lambda_Kramers}) need only be extended over nearest neighbors.

For long times $t \gg 1/\lambda_{1, \bm 0}^{\bm 0}$, the system dynamics is dominated by hopping within the $\alpha=0$ band and can be described by the single-band master equation (\ref{master_eqn_F_v2}) (see section \ref{sec:weak}).  In the tight-binding limit, $\lambda_{1, \bm 0}^{\bm 0}$ can be approximated by the splitting of the two lowest eigenvalues of the harmonic approximation to the minima of the periodic potential, i.e., $\lambda_{1, \bm 0}^{\bm 0} \sim \sum_j \partial ^2 V_{\bm 0}(\bm r)/\partial r_j^2|_{\bm r = \bm r_{\rm min}}/\gamma_j,$ where $\bm r _{\rm min}$ are the positions of the minima in the periodic potential.  

In the tight-binding limit, nearest neighbor hopping dominates and the summation in the master equation (\ref{master_eqn_F_v2}) need only be extended over nearest neighbors.  The physical interpretation of the master equation can then be made more explicit by assuming that the Wannier states in the lowest band are the Gaussian harmonic oscillator states of the potential minima.  With that assumption, the coefficients $p_{\bm n}(t)$ are non-negative according to Eq.\ (\ref{pncoeff}), provided that the Wannier states are chosen positive.  The coefficients  $p_{\bm n}(t)$ can then be interpreted as the probability of the particle being localized in the $\bm n$th potential well and the master equation (\ref{master_eqn_F_v2}) describes particle hopping between nearest-neighbor wells.  

\subsection{More General Potentials \label{sec:ext}}

In the tight-binding limit, the physical arguments used to derive the tilt dependence of the hopping rates $\kappa_{\alpha,\bm n}$ can be used to better understand Brownian particle dynamics in more general potentials.  For example, consider the external potential
\begin{equation}
V(\bm r) = V_{\bm 0}(\bm r)+W(\bm r),
\label{gen_pot}
\end{equation}
where $V_{\bm 0}(\bm r)$ is the periodic potential defined previously and the potential $W(\bm r)$ is slowly varying on the scale of $\bm a$.  In this case, the coupling matrix $\sigma_{\alpha,\alpha',\bm n,\bm n'}$ for the potential (\ref{gen_pot}) can be approximated by taking the first term in the Taylor expansion of $W(\bm r)$ around the position $\bm A \bm n$ of the appropriate Wannier state, i.e.,
\begin{eqnarray}
\lefteqn{ \frac{1}{D}\int d{\bm r} \ e^{V_0(\bm r)/\Theta} e^{W(\bm r)/2\Theta} w_{\alpha,\bm n}(\bm r) w_{\alpha',\bm n',}(\bm r) }\nonumber \\
& & \quad \quad \quad \quad \quad \quad \quad \quad \quad\approx e^{W(\bm A\bm n)/2\Theta} \delta_{\alpha\alpha'}\delta_{\bm n\bm n'}.
\end{eqnarray}
Approximating the operator ${\cal H}$ by ${\cal H}_0$ yields
\begin{equation}
\sigma_{\alpha,\alpha',\bm n,\bm n',} \approx e^{W(\bm A \bm n')/2\Theta}e^{-W(\bm A \bm n)/2\Theta} \kappa_{\alpha,\bm n-\bm n '}^{\bm 0} \delta_{\alpha \alpha'},
\label{sigma_gen_pot}
\end{equation}
and considering only nearest-neighbor hopping, Eq.\ (\ref{sigma_gen_pot}) can be approximated as
\begin{equation}
\sigma_{\alpha,\alpha',\bm n,\bm n'} \approx  e^{-\nabla W(\bm A \bm n)\cdot \bm A (\bm n - \bm n ' )/2\Theta} \kappa_{\alpha,\bm n-\bm n'}^{\bm 0} \delta_{\alpha\alpha'}.
\label{sigma_genV}
\end{equation}
To lowest order, $\nabla W(\bm A \bm n)$ is independent of $\bm n$ and the dynamics are described using the tilted periodic potential already discussed.  However, to higher orders Eq.\ (\ref{sigma_genV}) provides access to the dynamics for systems with external potentials of the form of Eq.\ (\ref{gen_pot}).

\section{Macroscopic Observables \label{sec:prop}}

Macroscopic properties of the system can be calculated directly from the probability density $P(\bm r,t)$ or by using either the untilted Bloch eigenfunction basis or the Wannier state basis.  We focus on the weak-tilting regime defined by Eq.\ (\ref{criterion}) for long times ($t \gg 1/\lambda_{1,\bm 0}^{\bm 0}$).  In this case the Wannier state basis is useful and the system is well described by the single-band master equation (\ref{master_eqn_F_v2}).  In the tight-binding limit the hopping rates are given by Eq.\ (\ref{kappa_f}).

\subsection{Drift}

The mean position can be calculated in the Wannier basis as
\begin{equation}
\langle {\bm r} \rangle (t) = \int  d{\bm r} \ {\bm r} P({\bm r},t) =  \sum _{\alpha,\bm n} p_{\alpha,\bm n}(t) \left({\bm R}_{\alpha}+{\bm A}  {\bm n} \xi_{\alpha}\right), 
\label{mean_pos}
\end{equation}
where
\begin{equation}
{\bm R}_{\alpha} = \frac{1}{D}\int d{\bm r} \ {\bm r} w_{\alpha,{\bm 0}}({\bm r}).
\end{equation}

In the weak-tilting regime retaining only the $\alpha=0$ band, the time derivative of Eq.\ (\ref{mean_pos}) can be found by inserting the master equation (\ref{master_eqn_F_v2}).  This yields
\begin{eqnarray}
\frac{d\langle {\bm r}\rangle(t) }{dt} & = & {\cal N} \sum_{\bm n} {\bm A}{\bm n}\kappa_{\bm n} \\
& = & -i {\cal N}\left.\nabla_{\bm k}  \lambda_{{\bm k}}\right|_{{\bm k}={\bm 0}} \\
& = &  {\cal N} \left.\nabla_{\bm k}  {\rm Im}(\lambda_{{\bm k}})\right|_{{\bm k}={\bm 0}} ,
\label{vel}
\end{eqnarray}
where $\cal N$ is the normalization.  Integrating Eq.\ (\ref{vel}) gives
\begin{equation}
\langle {\bm r}\rangle (t)=  \langle {\bm r}\rangle (0)+ {\cal N} t \left.\nabla_{\bm k}  {\rm Im} (\lambda_{{\bm k}}) \right|_{{\bm k}={\bm 0}},
\label{pos}
\end{equation}
and the drift is
\begin{equation}
\langle \dot{\bm r} \rangle (t)_{\rm st} = \lim_{t\rightarrow\infty} \frac{\langle \bm r \rangle (t)}{t} = {\cal N} \left.\nabla_{\bm k}  {\rm Im}(\lambda_{{\bm k}})\right|_{{\bm k}={\bm 0}}.
\label{drift}
\end{equation}
Equation (\ref{drift}) shows that the drift is proportional to the gradient of the imaginary part of the lowest Bloch band eigenvalues \cite{Jung95}.  For the untilted case, $\nabla_{\bf k} \lambda_{{\bm k}}^{\bm 0}$ vanishes at $\bm k = \bm 0$ [see Eqs.\ (\ref{kappa0}) and (\ref{kappa_sym})].  Therefore, $\langle {\bm r}_{\bm 0}\rangle (t)= \langle {\bm r}_{\bm 0}\rangle(0)$ and the drift vanishes.  This is a consequence of detailed balance for a periodic potential.

In the tight-binding limit, the tilt dependence of the hopping rates and the eigenvalues $\lambda_{\bm k}$ are known analytically (see section \ref{sec:Kramers}) and we find that
\begin{equation}
\frac{d\langle {\bm r}\rangle(t) }{dt}  = {\cal N} \sum_{\bm n} {\bm A}{\bm n}\kappa_{\bm n}^{\bm 0} e^{\bm f \cdot \bm A \bm n /2\Theta},
\label{drift_tb}
\end{equation}
where the sum is taken over nearest neighbors.  Considering only one dimension labeled $x$, the net rate of nearest neighbour hopping is
\begin{equation}
\kappa_{1}-\kappa_{-1} = -2\kappa_{1}^{0} \sinh \left(\frac{f_xa_x}{2\Theta}\right),
\label{net_rate_1d}
\end{equation}
and the eigenvalues of ${\cal L}$ in the $\alpha=0$ band are approximately given by
\begin{equation}
\lambda_{k_x} = 4\kappa_{1}^{0} \sin \left(\frac{k_x a_x}{2}\right) \sin \left(\frac{k_x a_x}{2}+i\frac{f_x a_x}{2\Theta}\right). 
\label{lambda_1d}
\end{equation}
Differentiating Eq.\ (\ref{lambda_1d}), the drift is
\begin{equation}
\langle \dot{x} \rangle(t) _{\rm st} = 2{\cal N}\kappa_{1}^0 a_x \sinh \left( \frac{f_x a_x}{2\Theta}\right).
\label{drift_1d}
\end{equation}
Equation (\ref{drift_1d}) is consistent with previous one-dimensional treatments \cite{Reimann01, Reimann02}.  In two or more dimensions, Eq.\ (\ref{drift_tb}) shows that a tilt in one dimension can induce current in another, provided that the potential is not separable \cite{Challis12}.

\subsection{Diffusion}

The second moment can be calulated to be
\begin{eqnarray}
\langle {\bm r}^2 \rangle(t) & = & \int  d{\bm r} \ {\bm r}^2 P({\bm r},t) \\
& = & \sum _{\alpha,\bm n} p_{\alpha,\bm n}(t) \left[ S_{\alpha}+2 {\bm A}  {\bm n} \cdot {\bm R}_{\alpha}+({\bm A \bm n})^2 \xi_{\alpha}\right],
\label{second_mom}
\end{eqnarray}
where
\begin{equation}
S_{\alpha} = \frac{1}{D}\int d{\bm r} \ {\bm r}^2 w_{\alpha, {\bm 0}} ({\bm r}).
\end{equation}

In the weak-tilting regime retaining only the $\alpha=0$ band, the time derivative of Eq.\ (\ref{second_mom}) can be found to be
\begin{eqnarray}
\frac{d\langle {\bm r}^2 \rangle }{dt}(t) & = & \frac{1}{{\cal N}}\frac{d\langle {\bm r} \rangle ^2(t)}{dt}+{\cal N}\sum_{\bm n}({\bm A \bm n})^2 \kappa_{\bm n} \\
& = & \frac{1}{{\cal N}}\frac{d\langle {\bm r}\rangle ^2(t)}{dt}+{\cal N}\left. \nabla^2_{\bm k} \lambda_{\bm k}\right|_{{\bm k}={\bm 0}} \\
& = & \frac{1}{{\cal N}}\frac{d\langle {\bm r} \rangle ^2(t)}{dt}+{\cal N}\left. \nabla^2_{\bm k} {\rm Re}(\lambda_{\bm k})\right|_{{\bm k}={\bm 0}} .
\end{eqnarray}
It is convenient to define the variance
\begin{equation}
{\rm var} (t)= \langle {\bm r}^2 \rangle (t)-\frac{1}{{\cal N}}\langle {\bm r}\rangle ^2(t),
\end{equation}
which evolves according to
\begin{equation}
\frac{d {\rm var}(t)}{dt} = {\cal N}\left. \nabla^2_{\bm k} {\rm Re}(\lambda_{\bm k}) \right|_{{\bm k}={\bm 0}}.
\label{rate_var}
\end{equation}
Integrating Eq.\ (\ref{rate_var}) gives
\begin{equation}
{\rm var}(t) = {\rm var}(0)+{\cal N} t \left. \nabla^2_{\bm k} {\rm Re} (\lambda_{\bm k})\right|_{{\bm k}={\bm 0}},
\label{diff}
\end{equation}
and the diffusion coefficient
\begin{equation}
{\cal D} = \lim_{t\rightarrow\infty} \frac{{\rm var}(t)}{2t} = \frac{1}{2}{\cal N} \left. \nabla^2_{\bm k} {\rm Re} (\lambda_{\bm k})\right|_{{\bm k}={\bm 0}}.
\label{diffusion}
\end{equation}
Equation (\ref{diffusion}) shows that the diffusion is proportional to the curvature of the real part of the lowest Bloch band eigenvalues \cite{Jung95}.  In the tight-binding limit, the diffusion in one dimension is
\begin{equation}
{\cal D} = {\cal N}\kappa_{1}^0 a_x^2 \cosh \left(\frac{f_x a_x}{2\Theta} \right).
\end{equation}

\section{Conclusion \label{sec:conc}}

We have presented a systematic transformation from the continuous Smoluchowski equation to a discrete master equation using a Wannier basis expansion analogous to the tight-binding model of a quantum particle in a periodic potential.  The discrete master equation can be interpreted in terms of intraband and interband hopping and, in the regime of weak tilting and long times, reduces to a single-band master equation consistent with the idea that the system relaxes rapidly within potential wells and is then dominated by comparatively slow hopping transitions between wells \cite{vanKampen77}. The regime of validity is equivalent to Kramers' regime for a bistable potential \cite{vanKampen78, Caroli79}.

In the tight-binding limit we have derived simple analytic expressions for the tilt dependence of the hopping rates and eigenvalues of the lowest Bloch band.  These expressions have been used to determine the tilt dependence of the drift and diffusion in the long-time limit.  When the periodic potential has a single dominant maximum per period the hopping rates can be approximated by Karmers' relation \cite{Caroli80, Hanggi90}, although our rates for $\bm f \neq \bm 0$ are restricted to weak tilting so do not depend on the relative position of the potential maxima between minima \cite{Fisher99, VandenBroeck12}.  The tilt dependence of the ratio between forward and backward hopping rates takes the form of a generalized detailed balance condition that is consistent with nonequilibrium fluctuation theorems \cite{Astumian07, Crooks99}. 

In this paper we have demonstrated that the tight-binding approach is a valuable tool for deriving analytic solutions to Fokker-Planck equations with periodic potentials.  This is particularly valuable for non-separable multidimensional potentials.  The implications of this work for energy transfer in molecular motors have been highlighted elsewhere \cite{Challis12}.

\appendix

\section{Orthonormality of Eigenfunctions \label{app:orthonorm}}

The self-adjoint operator ${\cal H}$ [see Eq.\ (\ref{Herm_op})] has the eigenvalue problem 
\begin{equation}
{\cal H}\psi_{\bm k}({\bm r}) = -\lambda_{\bm k} \psi_{\bm k}({\bm r}),
\label{Hestates}
\end{equation}
where the eigenfunctions are
\begin{equation}
\psi_{\bm k}({\bm r}) = e^{V({\bm r})/2 \Theta}\phi_{\bm k}({\bm r}),
\label{phitopsi}
\end{equation}
and $\lambda_{\bm k}$ and $\phi_{\bm k}({\bm r})$ are, respectively, the eigenvalues and eigenfunctions of the operator $\cal L$ [see Eq.\ (\ref{Lestates})].  Equation (\ref{phitopsi}) shows that, when the eigenfunctions $\phi_{\bm k}({\bm r})$ are chosen with periodic boundary conditions, the eigenfunctions $\psi_{\bm k}({\bm r})$ are not periodic.  This means that for the tilted periodic potential the operator ${\cal H}$ is in general not Hermitian, the eigenvalues $\lambda_{\bm k}$ may be imaginary, and the eigenfunctions $\psi_{\bm k}({\bm r})$ and $\phi_{\bm k}({\bm r})$ do not necessarily form a complete orthonormal basis \cite{Risken89}.  

For the untilted periodic potential with ${\bm f}={\bm 0}$, the operator ${\cal L}_{\bm 0}$ can be transformed to the self-adjoint operator ${\cal H}_{\bm 0}$ with eigenfunctions 
\begin{equation}
\psi^{\bm 0}_{\bm k}({\bm r}) = e^{V_{\bm 0}({\bm r})/2 \Theta}\phi^{\bm 0}_{\bm k}({\bm r}).
\label{phitopsi0}
\end{equation}
Equation (\ref{phitopsi0}) shows that both the eigenfunctions $\phi^{\bm 0}_{\bm k}({\bm r})$  and $\psi^{\bm 0}_{\bm k}({\bm r})$ can simulaneously have periodic boundary conditions.  For periodic boundary conditions with a period given by ${\bm N \bm a}$, the operator ${\cal H}_{\bm 0}$ is Hermitian.  This means that the eigenvalues $\lambda_{\bm k}^{\bm 0}$ are real and exist in continuous bands separated by finite forbidden band gaps \cite{Kittel04}.  The eigenfunctions $\psi^{\bm 0}_{\bm k}({\bm r})$ form a complete orthonormal basis \cite{Risken89}.  For an infinite spatial extent, the eigenfunctions $\psi^{\bm 0}_{\bm k}({\bm r})$ satisfy the orthonormality relation
\begin{equation}
\int d{\bm r} \ \psi_{\bm k}^{\bm 0*}({\bm r}) \psi^{\bm 0}_{{\bm k}'}({\bm r}) = \delta({\bm k}-{\bm k}'),
\end{equation} 
and the completeness relation
\begin{equation}
 \int d{\bm k} \ \psi_{\bm k}^{\bm 0 *} ({\bm r}) \psi^{\bm 0}_{\bm k} ({\bm r}') = \delta ({\bm r}-{\bm r}').
\end{equation}
In the ${\cal L}_{\bm 0}$ operator basis, the othonormality relation becomes
\begin{equation}
\int d{\bm r} \ e^{V_{\bm 0}({\bm r})/\Theta} \phi_{\bm k}^{\bm 0 *}({\bm r}) \phi^{\bm 0}_{{\bm k}'}({\bm r}) = \delta({\bm k}-{\bm k}'),
\label{completev0}
\end{equation}
and the completeness relation becomes
\begin{equation}
e^{ V_{\bm 0}({\bm r}) /\Theta} \int d{\bm k} \   \phi_{\bm k}^{\bm 0 *} ({\bm r}) \phi^{\bm 0}_{\bm k} ({\bm r}') = \delta ({\bm r}-{\bm r}').
\end{equation}

\section{Positivity of Eigenvalues \label{app:pose}}

Following the methodology presented in Ref.\ \cite{Risken89}, the operator ${\cal L}_{\bm 0}$ can be written as
\begin{equation}
{\cal L}_{\bm 0} = \sum_{j} \frac{\Theta}{\gamma_j}\frac{\partial}{\partial r_j} e^{-V_{\bm 0}({\bm r})/\Theta} \frac{\partial}{\partial r_j} e^{V_{\bm 0}({\bm r})/\Theta}.
\label{L0form}
\end{equation}
Using Eq.\ (\ref{Lestates}) with ${\bm f}= \bm 0$ and Eq.\ (\ref{completev0}),
\begin{eqnarray}
\lefteqn{\int d{\bm r} \ e^{V_{\bm 0}({\bm r})/\Theta} \phi^{\bm 0 *}_{{\bm k}} ({\bm r}){\cal L}_{\bm 0} \phi^{\bm 0}_{{\bm k}'}({\bm r})=  -\lambda_{{\bm k}}^{\bm 0} \delta ({\bm k}-{\bm k}')} \label{ekint} \\
& & =  - \int d{\bm r} \ e^{-V_{\bm 0}({\bm r})/\Theta} \sum_{j} \frac{\Theta}{\gamma_j} \left| \frac{\partial}{\partial r_j} e^{V_{\bm 0}({\bm r})/\Theta} \phi^{\bm 0}_{{\bm k}}({\bm r})\right|^2 ,
\label{ekint2}
\end{eqnarray}
where the left-hand side of Eq.\ (\ref{ekint}) has been integrated by parts using Eq.\ (\ref{L0form}).  Equation (\ref{ekint2}) is less than or equal to zero so the eigenvalues $\lambda_{{\bm k}}^{\bm 0}$ are greater than or equal to zero.  In particular, the ground state is
\begin{equation}
\phi^{\bm 0}_{{\bm 0}}({\bm r}) \propto e^{-V_{\bm 0}({\bm r})/\Theta},
\label{phi0}
\end{equation}
with the eigenvalue $\lambda^{\bm 0}_{\bm 0}=0.$

\section{Eigenvalues for the Weakly-Tilted Periodic Potential \label{app:evalues_tilt}}

The eigenfunctions $\phi_{\alpha,\bm k}({\bm r})$ of the tilted periodic potential can be expanded in the Wannier basis as
\begin{equation}
\phi_{\alpha,\bm k}({\bm r}) = \sum_{\alpha',{\bm n}'} \Phi_{\alpha,\alpha',{\bm k},{\bm n}'} w_{\alpha',{\bm n}'}({\bm r}).
\label{phiexpwan}
\end{equation}
Substituting expansion (\ref{phiexpwan}) into Eq.\ (\ref{Lestates}), multiplying by $e^{V_{\bm 0}({\bm r})/\Theta} w_{\alpha,\bm n}({\bm r})$, and integrating gives
\begin{equation}
-\lambda_{\alpha,\bm k}\Phi_{\alpha,\alpha,{\bm k},\bm n} = \sum_{\alpha',{\bm n}'} \Phi_{\alpha,\alpha',{\bm k},{\bm n}'} \sigma_{\alpha,\alpha',{\bm n},{\bm n}'}.
\label{eeqn}
\end{equation}
Using the Bloch form for $\phi_{\alpha,\bm k}({\bm r})$, it can be shown that
\begin{equation}
\Phi_{\alpha,\alpha,{\bm k},\bm n}=\Phi_{\alpha,\alpha',{\bm k},{\bm 0}}e^{i {\bm k} \cdot {\bm A \bm n}} .
\label{Phin}
\end{equation}
Substituting Eq.\ (\ref{Phin}) into Eq.\ (\ref{eeqn}) and neglecting interband coupling yields 
\begin{equation}
\lambda_{\alpha,\bm k} = -\sum_{\bm n}\kappa_{\alpha,\bm n} e^{-i \bm k \cdot \bm A\bm n},
\end{equation}
which can be inverted to derive Eq.\ (\ref{kappa}).  

\section{Normalization of the lowest untilted Bloch band \label{app:norm}}

The normalization of the probability density $P({\bm r},t)$ can be calculated in the Wannier state basis as
\begin{eqnarray}
{\cal N} (t) & = & \int d{\bm r} \ P({\bm r},t) = \sum_{\alpha,\bm n}   p_{\alpha,\bm n}(t) \xi_{\alpha} ,
\label{norm_wbasis}
\end{eqnarray}
where
\begin{equation}
\xi_{\alpha} =\frac{1}{D} \int d{\bm r} \ w_{\alpha,{\bm 0}} ({\bm r}).
\end{equation}
In the $\bm f = \bm 0$ case, the normalization can be determined by taking the sum over $\bm n$ of Eq.\ (\ref{pneqnv2}) and integrating to give
\begin{equation}
{\cal N}_{\bm 0}(t) = \sum_{\alpha,\bm n}  p^{\bm 0}_{\alpha,\bm n}(0) e^{-\lambda_{\alpha,{\bm 0}}^{\bm 0} t} \xi_{\alpha} .
\label{Nsol}
\end{equation}
Equation (\ref{Nsol}) shows that the band $\alpha$ decays in time with a rate given by $\lambda^{\bm 0}_{\alpha,{\bm 0}} \geq 0$ (see Appendix \ref{app:pose}).  Only the $\alpha=0$ band with $\lambda_{0,{\bm 0}}^{\bm 0}=0$ is stable.  This is consistent with the separation of time-scales argument presented in section \ref{sec:Bloch} [see Eq.\ (\ref{diag_sol_0})].

For the tilted periodic potential, taking the time derivative of Eq.\ (\ref{norm_wbasis}) and inserting the master equation (\ref{pneqnv3}), the time derivative of the normalization can be found to be
\begin{equation}
\frac{d {\cal N}(t)}{dt} = \sum_{\alpha,\bm n,\bm n '}\left[   \kappa_{\alpha,\bm n '}^{\bm 0}p_{\alpha,\bm n}(t)  +\sum_{\alpha'}\Delta _{\alpha,\alpha',\bm n'}p_{\alpha',\bm n}(t) \right]  \xi_{\alpha}.
\label{norm_f}
\end{equation}
The contribution from the lowest $\alpha=0$ band is given by the $\alpha=0$ contribution to the sum in Eq.\ (\ref{norm_f}).  The first term is due to the periodic potential and vanishes because $\sum_{\bm n}\kappa_{0,\bm n}^{\bm 0}=-\lambda^{\bm 0}_{0, {\bm 0}}=0$ [see Eq.\ (\ref{kappa0}) and Appendix \ref{app:pose}].  The second term is proportional to $\sum_{\bm n} \Delta _{0,\alpha,\bm n} =  \nu_{0, \alpha, \bm 0} $ and
\begin{eqnarray}
\nu_{0, \alpha, \bm 0}  & \propto &  \int d{\bf r} \ e^{V_{\bm 0}({\bm r})/\Theta} \phi_{0, {\bm 0}}^{\bm 0*}({\bm r})  \sum_{j} \frac{f_j}{\gamma_j} \frac{\partial}{\partial r_j} \phi_{\alpha,{\bm 0}}^{\bm 0}  ({\bm r}) \\
& =&   \int d{\bf r} \ \sum_{j}  \frac{f_j}{\gamma_j} \frac{\partial}{\partial r_j}  \phi_{\alpha,{\bm 0}}^{\bm 0}  ({\bm r}) \label{usephi} \\
& = & 0, 
\end{eqnarray}
where we have used that the untilted Bloch ground state $\phi_{0,{\bm 0}}^{\bm 0}({\bm r})\propto e^{-V_{\bm 0}(\bm r)/\Theta}$ [see Eq.\ (\ref{phi0})] is periodic at the boundary.  This shows that the untilted $\alpha=0$ Bloch band is stable when evolved by the operator $\cal L$, even for a finite tilt.  

\section{Validity of the Tight-Binding Limit \label{app:tight_binding}}

In the tight-binding limit, the shape of the Wannier states depends on the curvature of the periodic potential at its extrema and in the vicinity of these extrema the linear potential is a small perturbation.  To derive the regime of validity for the tight-binding limit, the periodic potential in the vicinity of its extrema at $\bm r = \bm r_{\rm ext}$ can be approximated by the harmonic form
\begin{eqnarray}
V_{\bm 0}(\bm r) \approx V_{\bm 0}(\bm  r_{\rm ext})+  \frac{1}{2} \sum_j ( r_j-r_{{\rm ext} j})^2\left. \frac{\partial^2 V_{\bm 0}(\bm r')}{\partial r'^2_j}\right|_{\bm r' = \bm r_{\rm ext}}.
\label{pot_harm}
\end{eqnarray}
The components of the potential $U^{\bm 0}(\bm r)$ of the Hermitian operator ${\cal H}_{\bm 0}$ [see Eq.\ (\ref{Upot})] can then be approximated by
\begin{equation}
U^{\bm 0}_j (\bm r) \approx \left[\frac{1}{4\Theta} (r_j-r_{{\rm ext} j})^2 -\frac{1}{2}\right] \left.\frac{\partial^2 V_{\bm 0}(\bm r')}{\partial r_j'^2}\right|_{\bm r' = \bm r_{\rm ext}}.
\end{equation}
With the addition of the linear potential, the components of the potential $U(\bm r)$ become
\begin{equation}
U_j (\bm r) \approx \left[\frac{1}{4\Theta} \left(r_j-\bar{r}_{{\rm ext} j}\right)^2 -\frac{1}{2} \right] \left.\frac{\partial^2 V_{\bm 0}(\bm r')}{\partial r_j'^2}\right|_{\bm r' = \bm r_{\rm ext}},
\label{potU_harm}
\end{equation}
where the tilt shifts the extrema from $\bm r = \bm r_{\rm ext}$ to $\bm r =  \bar{\bm r}_{\rm ext}$ with
\begin{equation}
\bar{r}_{{\rm ext}j} = {r}_{{\rm ext}j} +\frac{f_j}{\left.\partial^2 V_{\bm 0}(\bm r) / \partial r_j^2\right|_{\bm r=\bm r_{\rm ext}}}.
\end{equation}
The tight-binding limit corresponds to the regime where the shift $|\bar{r}_{{\rm ext}j}-{r}_{{\rm ext}j}|$ is significantly smaller than the period $a_j$ of the periodic potential, i.e.,
\begin{equation}
|f_j| \ll  a_j \left| \frac{\partial^2 V_{\bm 0}(\bm r)}{\partial r_j^2}\right|_{\bm r=\bm r_{\rm ext}}.
\label{validity_tb}
\end{equation}

\end{document}